 \documentclass[final,5p,times,twocolumn,authoryear]{elsarticle}
 
\biboptions{numbers,sort&compress}
\bibpunct{[}{]}{,}{n}{}{;}

\usepackage{amsfonts,amsmath,amssymb}
\usepackage{lipsum}
\usepackage{xcolor}
\usepackage{siunitx}
\usepackage{hyperref}
\usepackage{booktabs}
\usepackage[sort&compress]{natbib}

\journal{Physics Letters B}

\begin{document}

\begin{frontmatter}

\title{From DUNE Sensitivities to UV Models: Implications of Flavour Constraints}

\author[first]{Adriano Cherchiglia}
\affiliation[first]{organization={Centro de Ciências Naturais e Humanas, Universidade Federal do ABC},
            addressline={Avenida dos Estados, 5001}, 
            city={Santo André},
            postcode={09280-560}, 
            state={São Paulo},
            country={Brazil}}

\begin{abstract}
Future neutrino experiments, in particular DUNE, are expected to probe signals of new physics. These effects can be conveniently parametrized in terms of Wilson coefficients in the LEFT, with direct connection to non-standard interactions at production, propagation and detection in the QFT formalism. We revisit the apparent sensitivity of DUNE to semileptonic electron--tau interactions within the Standard Model Effective Field Theory. We show that the vector combination controlling the $\epsilon_{e\tau}$ NSI is precisely the combination entering $\tau\to e\omega$. Therefore, within the dimension-six vector subspace considered in this work, an operator cancellation that suppresses the tau-decay amplitude also suppresses the corresponding NSI. As a concrete example, we focus on the lepton-flavour-violating semileptonic coefficient $(C^{(1)}_{\ell q,1311})$. We find the bound $8.1\times10^{-3}\,{\rm TeV}^{-2}$ at $90\%$ confidence level, which is one order of magnitude stronger than the DUNE's expected sensitivity. In view of this scenario, we further study one illustrative minimal UV completion, showing that it faces even stronger constraints from other sources, making it very challenging to build UV scenarios with heavy mediators that could produce flavour-changing NSIs observable at DUNE. 
\end{abstract}


\end{frontmatter}

\section{Introduction}
\label{introduction}

The discovery of neutrino oscillations established that lepton flavour is not an exact symmetry of Nature and provided compelling evidence that the Standard Model (SM), in its minimal form, is incomplete. The next generation of long-baseline neutrino experiments, and in particular DUNE, will test the three-neutrino paradigm with unprecedented precision and will be sensitive to subleading effects from new physics in neutrino production, propagation, and detection~\cite{DUNE:2020ypp,DUNE:2020jqi,DUNE:2020fgq,Arguelles:2019xgp}. These effects are commonly parametrized in terms of non-standard interactions (NSI)~\cite{Wolfenstein:1977ue,Mikheyev:1985zog,Farzan:2017xzy,Esteban:2018ppq}. In a gauge-invariant quantum-field-theory description, however, neutrino NSI are connected to effective operators that also modify charged-lepton and hadronic processes. The Standard Model Effective Field Theory (SMEFT) therefore provides a framework in which projected DUNE sensitivities can be consistently confronted with existing laboratory constraints~\cite{Bergmann:1999pk,Antusch:2008tz,Gavela:2008ra,Meloni:2009cg,Altmannshofer:2018xyo,Falkowski:2019kfn,Falkowski:2019xoe,Babu:2019mfe,Bischer:2019ttk,Davidson:2019iqh,Terol-Calvo:2019vck,Babu:2020nna,Du:2020dwr,Falkowski:2021bkq,Du:2021rdg,Breso-Pla:2023tnz,Coloma:2024ict,Kopp:2024yvh,Kopp:2025ffx,Freitas:2025bgg,Gonzalez-Alonso:2026sgl}.

A recent comprehensive study of DUNE sensitivities to SMEFT Wilson coefficients found that present bounds exceed the projected experimental reach in many directions~\cite{Kopp:2025ffx}. A potentially interesting exception involves semileptonic lepton-flavour-violating coefficients with electron and tau flavour indices. The viability of this apparent window is nevertheless unclear. Gauge invariance relates the neutrino interactions relevant for matter propagation to charged-lepton-flavour-violating semileptonic processes. Exclusive modes such as
\begin{equation}
    \tau \to e\eta,\qquad
    \tau \to e\eta',\qquad
    \tau \to e\omega,\qquad
    \tau \to e\rho
\end{equation}
probe combinations of the same SMEFT Wilson coefficients. 

In this work, we show that the neutral current NSI $\epsilon_{e \tau}$ to be probed by DUNE shares the same direction in the dimension-six SMEFT WC space as the $\tau\to\omega e$ amplitude. Hence, suppressing $\tau\to\omega e$ also suppresses the DUNE sensitivity to this NSI. The remaining tau-meson modes constrain complementary directions, which delimit the possible cancellation regions for the coefficient $C_{\ell q,1311}^{(1)}$. This provides an exclusive-process interpretation of the corresponding global-fit constraint found in~\cite{Fernandez-Martinez:2024bxg}.

We further investigate if the tau decay bounds can be saturated by weakly coupled ultraviolet (UV) completions, relying on the systematic connection between heavy mediators and the SMEFT at dimension-six obtained using tree-level~\cite{deBlas:2017xtg} and one-loop dictionaries~\cite{Guedes:2023azv,Guedes:2024vuf}. Building on the automated matching strategy for long-baseline neutrino experiments developed in~\cite{Cherchiglia:2023aqp,Cherchiglia:2025ufn}, we analyze an illustrative minimal realization. We find that bounds from other sources are even stronger, which constrain the upper limit on $C_{\ell q,1311}^{(1)}$ before the tau-decay bounds can be saturated. 
 
Our work is organized as follows: in section \ref{sec:eft_tau} we discuss the connection between the NSI direction probed by DUNE and semileptonic tau decays in detail. In section \ref{sec:minimal_uv} we present a minimal UV model, showing the even stronger constraints it faces. We conclude in section \ref{sec:conclusions}.

\section{DUNE-sensitive SMEFT directions and LFV tau decays}
\label{sec:eft_tau}

We begin with an analysis of the semileptonic SMEFT directions relevant for the apparent $e$--$\tau$ window identified in~\cite{Kopp:2025ffx}. Throughout this section, the Wilson coefficients
are taken to be real, quoted in units of $\mathrm{TeV}^{-2}$, and they are
defined at the scale $\mu=2~\mathrm{TeV}$\footnote{When considering UV completions, we have chosen the masses to lie at this scale. Thus, for consistency, we will throughout the manuscript adopt this value.}. Within these conventions, the SMEFT Lagrangian with terms of dimension-six only is given by 
\begin{equation}
    \mathcal{L}_{\mathrm{SMEFT}}
    =
    \mathcal{L}_{\mathrm{SM}}
    +
    \sum_i C_i\,\mathcal{O}_i\;.
    \label{eq:smeft_lagrangian}
\end{equation}

We are particularly interested in the semileptonic SMEFT WC given by
\begin{align}
    \mathcal{O}_{\ell q}^{(1)\,prst}
    &=
    \bigl(\bar{\ell}_p\gamma_\mu\ell_r\bigr)
    \bigl(\bar{q}_s\gamma^\mu q_t\bigr),
    \label{eq:Olq1}
    \\
    \mathcal{O}_{\ell u}^{\,prst}
    &=
    \bigl(\bar{\ell}_p\gamma_\mu\ell_r\bigr)
    \bigl(\bar{u}_s\gamma^\mu u_t\bigr),
    \label{eq:Olu}
    \\
    \mathcal{O}_{\ell d}^{\,prst}
    &=
    \bigl(\bar{\ell}_p\gamma_\mu\ell_r\bigr)
    \bigl(\bar{d}_s\gamma^\mu d_t\bigr).
    \label{eq:Old}
\end{align}
It will prove useful to introduce the combinations
\begin{equation}
    C_V
    \equiv
    \frac{C_{\substack {\ell u \\  1311}}+C_{\substack {\ell d \\  1311}}}{2},
    \qquad
    C_A
    \equiv
    \frac{C_{\substack {\ell u \\  1311}}-C_{\substack {\ell d \\  1311}}}{2}.
    \label{eq:CV_CA_definitions}
\end{equation}

As we showed in \cite{Cherchiglia:2023aqp}, there is a connection between NSI at propagation and SMEFT WCs. For Earth-based neutrino experiments, such as DUNE, one obtains at tree-level matching, when only $C^{(1)}_{\substack {\ell q}}\,,\,C_{\substack {\ell u}},\,C_{\substack {\ell d}}$ are non-zero,
\begin{align}\label{eq:NSI_prop}
     \epsilon_{e\tau} = 
     &- 3v^{2}\left[C^{(1)}_{\substack {\ell q \\ 1311}}+C_{V}\right]\,,
 \end{align}
where, for simplicity, we adopted a diagonal CKM in the above formula. We recall that the presence of $\epsilon_{\alpha\beta}$ modifies the propagation of neutrinos through matter. In particular, if $\epsilon_{e\tau}$ ($\epsilon_{13}$) is non-null, it implies mainly on a modification of the number of events predicted in the (anti)electron appearance channels at DUNE. 

Electroweak gauge invariance implies that the same operators also induce
charged-lepton-flavour-violating tau decays with mesons containing only first generation quarks. In particular, we will consider the following modes
\begin{equation}
    \tau\to e\eta,
    \qquad
    \tau\to e\eta',
    \qquad
    \tau\to e\omega,
    \qquad
    \tau\to e\rho.
    \label{eq:tau_modes}
\end{equation}

The decay $\tau\to e\omega$ is particularly relevant because
the $\omega$ meson couples to the isoscalar light-quark vector
current. In the LEFT notation of~\cite{Fernandez-Martinez:2024bxg}, its branching ratio,
neglecting tensor operators and the electron mass, is given by
\begin{align}
\mathrm{BR}(\tau\to e\omega)
={}&
\frac{
G_F^2 f_\omega^2m_\omega^2
(m_\tau^2-m_\omega^2)
}{
8\pi\Gamma_\tau m_\tau
}
\left(
\frac{m_\tau^2}{m_\omega^2}
+1
-2\frac{m_\omega^2}{m_\tau^2}
\right)
\nonumber\\
&\times
\left|
c^{uV}_{e\tau L}
+c^{dV}_{e\tau L}
\right|^2 .
\end{align}
Within the semileptonic SMEFT WCs of equations \ref{eq:Olq1}-\ref{eq:Old},
the tree-level matching gives
\begin{equation}
\frac{2}{v^2}
\left(
c^{uV}_{e\tau L}+c^{dV}_{e\tau L}
\right)
=
C_{\ell q,1311}^{(1)}
+
\frac{
C_{\ell u,1311}+C_{\ell d,1311}
}{2}.
\end{equation}
Thus, the decay probes the combination
\begin{equation}
C_\omega
=
C_{\ell q,1311}^{(1)}+C_V.
\end{equation}

As can be seen, at tree-level matching, we obtain that the NSI coefficient $\epsilon_{e\tau}$ probes the same direction as the $ \tau\to e\omega$ mode. Therefore, within the dimension-six semileptonic SMEFT WC considered in this section, a cancellation that suppresses the $\tau\to\omega e$ amplitude also suppresses $\epsilon_{e\tau}$. Thus, not only can the individual DUNE bounds on $C_{\ell q}^{(1)}$, $C_{\ell u}$, and $C_{\ell d}$ not be disentangled from the $\tau\to\omega e$ bound, but the flat direction probed by $\epsilon_{e\tau}$ is the same as the one probed by the tau decay.

It is useful to notice that the pseudoscalar modes probe complementary combinations, breaking the flat direction. Actually, the dominant combination associated with $\tau\to\eta e$ is proportional to
\begin{equation}
    C_{\eta}
    \propto
    C_{\ell q,1311}^{(1)}-C_V,
    \label{eq:Ceta}
\end{equation}
with the precise $\eta$ and $\eta'$ expressions containing the corresponding hadronic decay constants and mixing parameters. Finally, the $\rho e$ mode provides an additional isovector constraint, which, however, is independent of the SMEFT WC $ C_{\ell q,1311}^{(1)}$. 

Since \texttt{smelli} v.2.4.3~\cite{Straub:2018kue,Aebischer:2018bkb,Aebischer:2018iyb}, which is the version used in this work, does not include the tau modes discussed (with the exception of $\tau\to e\rho$), we have built a one-sided Gaussian approximation to the
experimental upper limit, following the same approach adopted there. Explicitly, 
\begin{equation}
    \chi_X^2
    =
    \left[
      \frac{\mathrm{BR}(\tau\to eX)^{\mathrm{theo}}}
           {\mathrm{BR}^{\mathrm{exp}}_X/1.64485}
    \right]^2,
    \label{eq:tau_chi2}
\end{equation}
where $\mathrm{BR}(\tau\to eX)^{\mathrm{theo}}$ is the theoretical prediction for the mode $\tau\to e X$ and $\mathrm{BR}^{\mathrm{exp}}_{X}$ is the 90$\%$ C.L. upper limit listed in table \ref{tab:lfv_limits}.

\begin{table}[htbp]
  \centering
    \begin{tabular}{@{}lll@{}}
      \toprule
      Decay mode & Upper limit 90$\%$ C.L. & Ref. \\
      \midrule
      $\tau^-\to e^-\eta$
        & $9.2\times10^{-8}$
        & Belle~\cite{Belle:2007cio} \\
       $\tau^-\to e^-\eta'$
        & $1.6\times10^{-7}$
        & Belle~\cite{Belle:2007cio} \\
      \addlinespace
      $\tau^-\to e^-\omega$
        & $2.4\times10^{-8}$
        & Belle~\cite{Belle:2023ziz} \\
      $\tau^-\to e^-\rho^0$
        & $2.2\times10^{-8}$
        & Belle~\cite{Belle:2023ziz} \\
      \bottomrule
    \end{tabular}
    \caption{Experimental upper limits used in LFV $\tau$-meson likelihood.}
    \label{tab:lfv_limits}
\end{table}

The full $\chi^2_{\tau}$ from the tau modes is finally given by\footnote{When this likelihood is combined with the \texttt{smelli} likelihood, the $\rho$ mode is omitted.}
\begin{equation}
    \chi^2_{\tau}
    =
    \sum_{X=\eta,\eta',\omega, \rho}\chi_X^2,
    \label{eq:tau_total_chi2}
\end{equation}

In the following analysis we will consider the likelihood due only to the tau modes, since it provides stronger bounds. In the next section, when discussing a given UV model, we will include both the likelihood from \texttt{smelli} as well as our own implementation of $D^0$--$\bar D^0$ mixing. We first consider the case in which only $C^{(1)}_{\substack {\ell q \\ 1311}}$ is non-null. At 90$\%$ C.L., we obtain the bound 
\begin{align}
    \left|C^{(1)}_{\substack {\ell q \\ 1311}}\right|<8.1\times10^{-3}\;\rm{TeV}^{-2}\,.
\end{align}
This result lies well below the DUNE benchmark discussed in~\cite{Kopp:2025ffx}, which we extract as\footnote{We disregard the small correction between evolving the result quoted in  ~\cite{Kopp:2025ffx} at 1 TeV and ours at 2 TeV.}
\begin{align}
    \left|C^{(1)}_{\substack {\ell q \\  1311}}\right|\sim9\times10^{-2}\;\rm{TeV}^{-2}\,.
    \label{eq:dune_target}
\end{align}
The latter corresponds to $\epsilon_{e\tau}<0.016$, stronger than the bound found, for instance, in~\cite{Coloma:2023ixt}.

\begin{figure}[h!]
    \centering
    \includegraphics[scale=0.5]{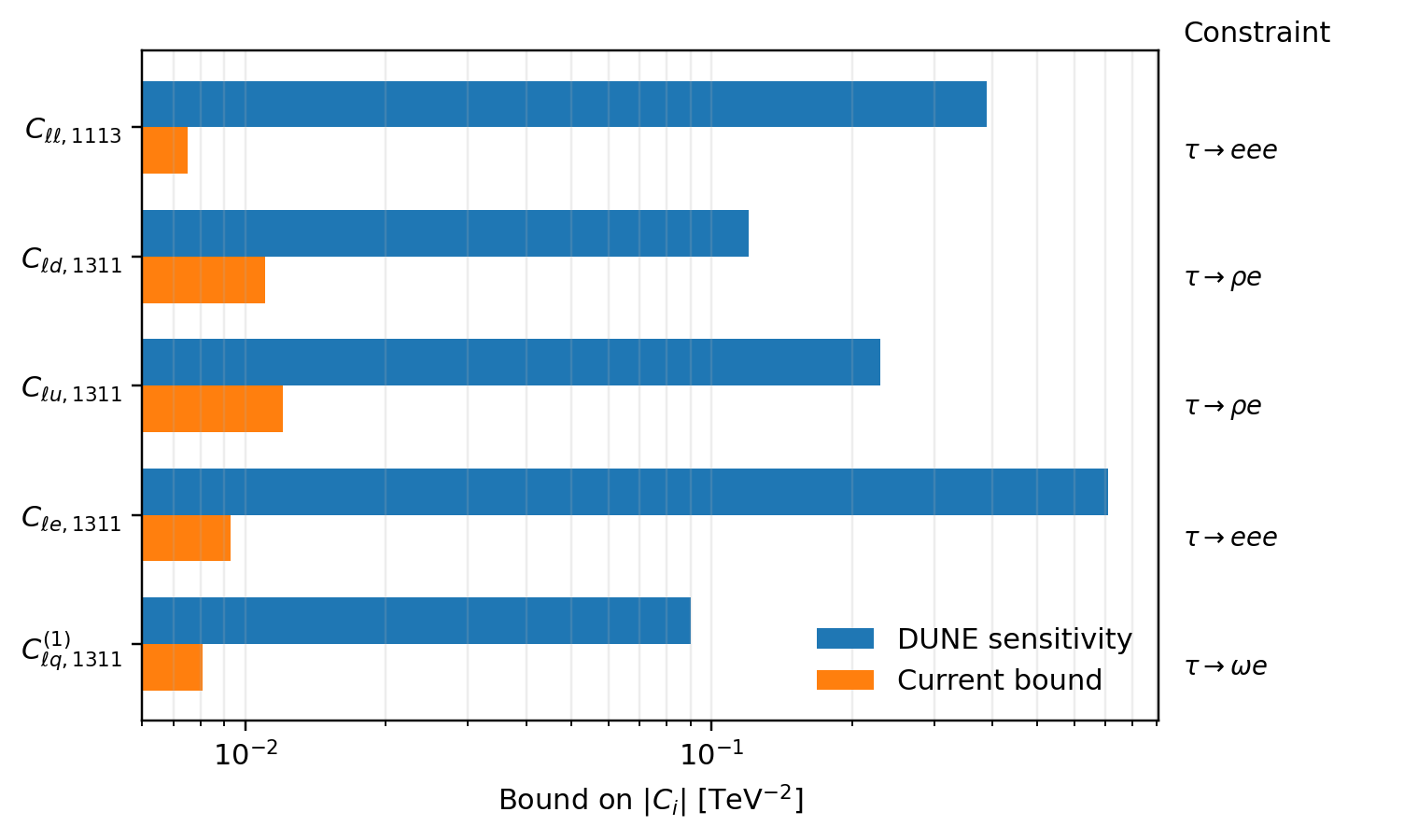}
    \caption{Comparison between the DUNE bounds extracted from \cite{Kopp:2025ffx} and bounds from semileptonic and leptonic tau decays.}
    \label{fig:dune_comparison}
\end{figure}

We performed a similar analysis for all the e-$\tau$ semileptonic and leptonic WC emphasized in \cite{Kopp:2025ffx}, whose results are shown in figure \ref{fig:dune_comparison}

It is clear from the figure that DUNE's capability to constrain the semileptonic SMEFT WC are one order of magnitude weaker than present bounds from tau decays. Even more importantly, since the DUNE sensitivity is extracted from bounds on $\epsilon_{e\tau}$, there is no easy way to evade the tau decay bounds as we showed earlier.

\begin{figure}[h!]
    \centering
    \includegraphics[scale=0.4]{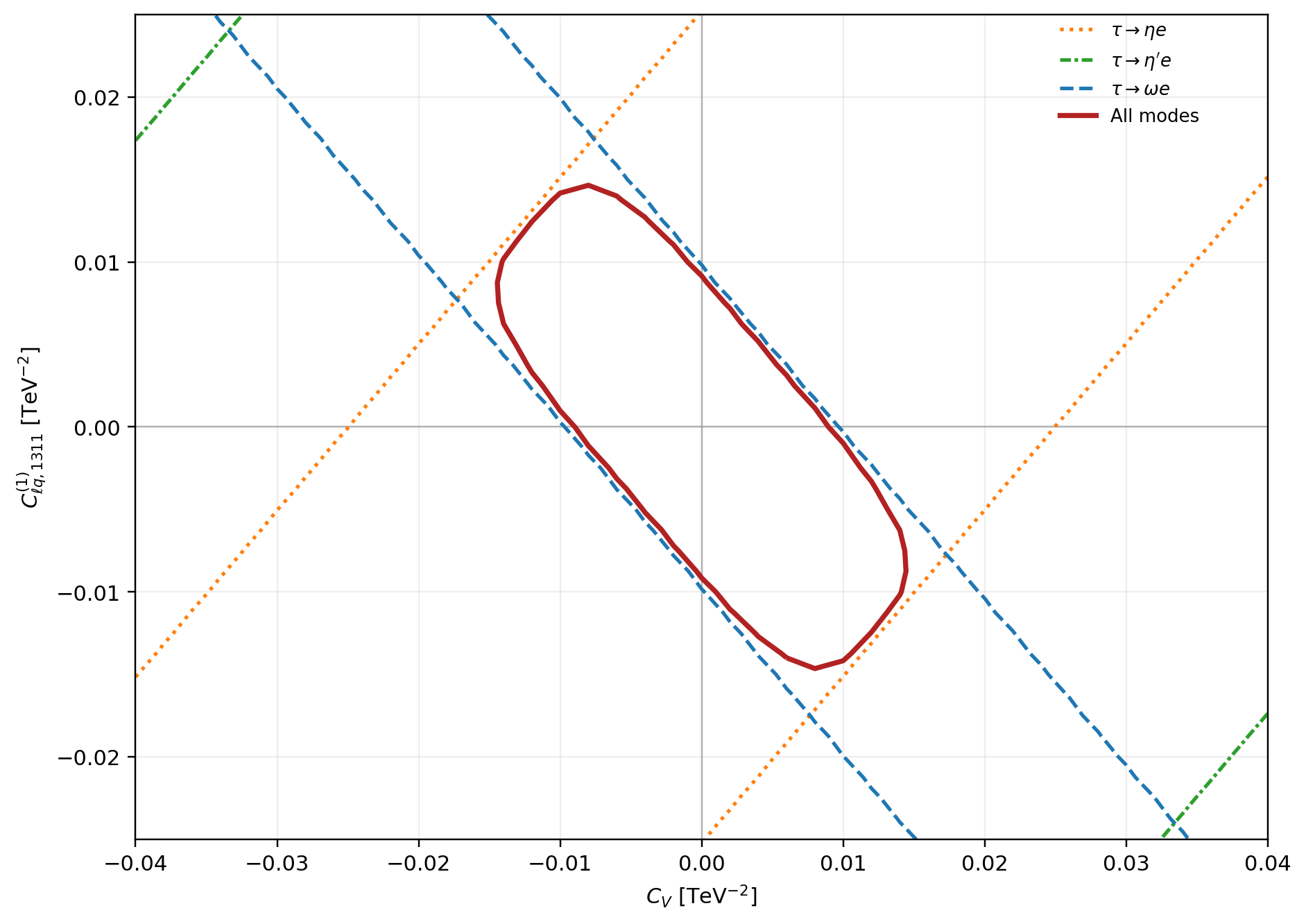}
    \caption{
      LFV tau-meson constraints in the
      $\bigl(C_{\ell q,1311}^{(1)},C_V\bigr)$ plane after profiling over
      $C_A$. The red contour shows the combined $90\%$ confidence region.
    }
    \label{fig:eft_tau_plane}
\end{figure}

We end this section by noting that the previous discussion assumes a single nonzero WC at a time. Nevertheless, ultraviolet completions generally generate several semileptonic operators simultaneously, and correlations among them may weaken bounds on an individual coefficient. To test whether such correlations may relax the bound on $C^{(1)}_{\substack {\ell q \\  1311}}$ given the restricted subspace of WCs given by equations \ref{eq:Olq1}-\ref{eq:Olu}, we perform a scan in the $(C_{\ell q,1311}^{(1)},C_V)$ plane, profiling over $C_A$. The results are shown in figure \ref{fig:eft_tau_plane}.

Since the $\tau \to \omega e$ and $\tau \to \eta e$ modes probe approximately orthogonal directions in the plane considered, one can extract an upper bound on $C^{(1)}_{\substack {\ell q \\ 1311}}$ which, after profiling over $C_V$ and $C_{A}$, is given by
\begin{equation}
    \left|C^{(1)}_{\substack {\ell q \\ 1311}}\right|
    \lesssim
    1.4\times10^{-2}\ \mathrm{TeV}^{-2}
    \qquad (90\%~\mathrm{C.L.}).
    \label{eq:profiled_clq_tau_bound}
\end{equation}
Our result is compatible with the bounds extracted from the global fit performed in~\cite{Fernandez-Martinez:2024bxg}, which indicates that $\tau\to\omega e$ and $\tau\to\eta e$ are the main observables able to constrain $C^{(1)}_{\substack {\ell q \\ 1311}}$.

\section{A minimal ultraviolet realization}
\label{sec:minimal_uv}

The analysis of Sec.~\ref{sec:eft_tau} treats the SMEFT WCs as independent parameters. In an ultraviolet (UV) completion, however, the coefficients are usually correlated functions of a smaller set of couplings. The purpose of this section is to determine whether the upper limit on $C^{(1)}_{\substack {\ell q \\ 1311}}$ from tau decays mode can be reached in a minimal UV realization.

In order to connect specific SMEFT WC to UV models, it is particularly useful to use dictionaries, where the matching between a general BSM Lagrangian to the SMEFT was done once and for all. Regarding dimension-six operators, the complete tree-level dictionary was derived in~\cite{deBlas:2017xtg}. This classification was applied explicitly to neutrino NSI in~\cite{Cherchiglia:2023aqp}. Regarding $\epsilon_{e\tau}$, it is directly connected to the WCs given in eqs.\ref{eq:Olq1}-\ref{eq:Old}, whose tree-level realization, restricting ourselves to isosinglets or isodoublets of scalars and/or fermions, is given by
\begin{align}
    \omega_1 &\sim (\mathbf{3},\mathbf{1})_{-1/3},
    &
    &\mathcal{O}_{\ell q}^{(1)},
    \label{eq:omega1_irrep}
    \\
    \Pi_7 &\sim (\mathbf{3},\mathbf{2})_{7/6},
    &
    &\mathcal{O}_{\ell u},
    \label{eq:Pi7_irrep}
    \\
    \Pi_1 &\sim (\mathbf{3},\mathbf{2})_{1/6},
    &
    &\mathcal{O}_{\ell d}.
    \label{eq:Pi1_irrep}
\end{align}
As can be seen, for the WCs of interest, only scalar leptoquarks are needed.

Since we have been focusing on $C^{(1)}_{\substack {\ell q \\ 1311}}$, we will consider the leptoquark $\omega_1$ as our minimal realization. We have checked that similar results are obtained for the other cases as well. Explicitly, we will consider the interactions
\begin{equation}
    \mathcal{L}_{\omega_1}
    \supset
    \lambda_e\,
    \omega_1^\dagger\,
    \overline{q_{L1}^{\,c}}\,
    i\sigma_2\ell_{L1}
    +
    \lambda_\tau\,
    \omega_1^\dagger\,
    \overline{q_{L1}^{\,c}}\,
    i\sigma_2\ell_{L3}
    +\mathrm{h.c.},
    \label{eq:omega1_lagrangian}
\end{equation}
where $q_{L1}$ is the first-generation quark doublet and $\ell_{L1,3}$ are the electron and tau lepton doublets. We take the couplings to be real, for simplicity.

We should emphasize that other couplings are allowed in general, including diquark interactions. Such interactions would generate baryon-number-violating operators and we set them to zero. This choice can be enforced, for instance, by assigning baryon number to the leptoquark. All other gauge-allowed $\omega_1$ couplings are likewise set to zero.

Integrating out $\omega_1$ at tree level gives~\cite{deBlas:2017xtg}
\begin{align}
   C^{(1)}_{\substack {\ell q \\ 1311}}
    =
    \frac{\lambda_e^\ast\lambda_\tau}
         {4M_{\omega_1}^{2}},
 \quad   C^{(3)}_{\substack {\ell q \\ 1311}}
    =
    -C^{(1)}_{\substack {\ell q \\ 1311}}.
    \label{eq:omega1_matching}
\end{align}
To evade current direct-search limits at the LHC~\cite{Allwicher:2022gkm}, we set $M_{\omega_1}=2~\mathrm{TeV}$. To good approximation, results for other masses can be obtained by rescaling.

We evaluate the model by matching it onto the SMEFT at the heavy scale,
evolving the Wilson coefficients to the relevant low-energy scales, and
constructing the likelihood
\begin{equation}
    \chi^2_{\mathrm{tot}}
    =
    \chi^2_{\mathrm{global}}
    +
    \chi^2_{D\text{-mix}}
    +
    \chi^2_{\tau\text{-mesons}}.
    \label{eq:uv_total_chi2}
\end{equation}
The first term is obtained from the global likelihood implemented in \texttt{smelli}, with the running and matching performed using \texttt{wilson} and the low-energy observables evaluated with \texttt{flavio}. The tau-meson contribution contains the modes discussed in Sec.~\ref{sec:eft_tau}. We also include the real and imaginary parts of the $D^0$--$\overline D{}^0$-mixing amplitude, following~\cite{DiLuzio:2018zxy}. We assume throughout the down-aligned basis for quarks flavors, in order to avoid the even stronger bounds from kaon physics. 

We should emphasize that although the target SMEFT WC is generated at tree level, the $D$-mixing contribution is a genuine consequence of the complete UV matching at one-loop. In particular, loop-induced four-quark coefficients, together with flavour rotations and renormalization-group evolution, correlate the semileptonic WC with neutral-meson mixing. For this purpose we employ the automated one-loop dictionary \texttt{SOLD}~\cite{Guedes:2023azv,Guedes:2024vuf}.

\begin{figure*}[h!]
    \centering
    \includegraphics[scale=0.5]
        {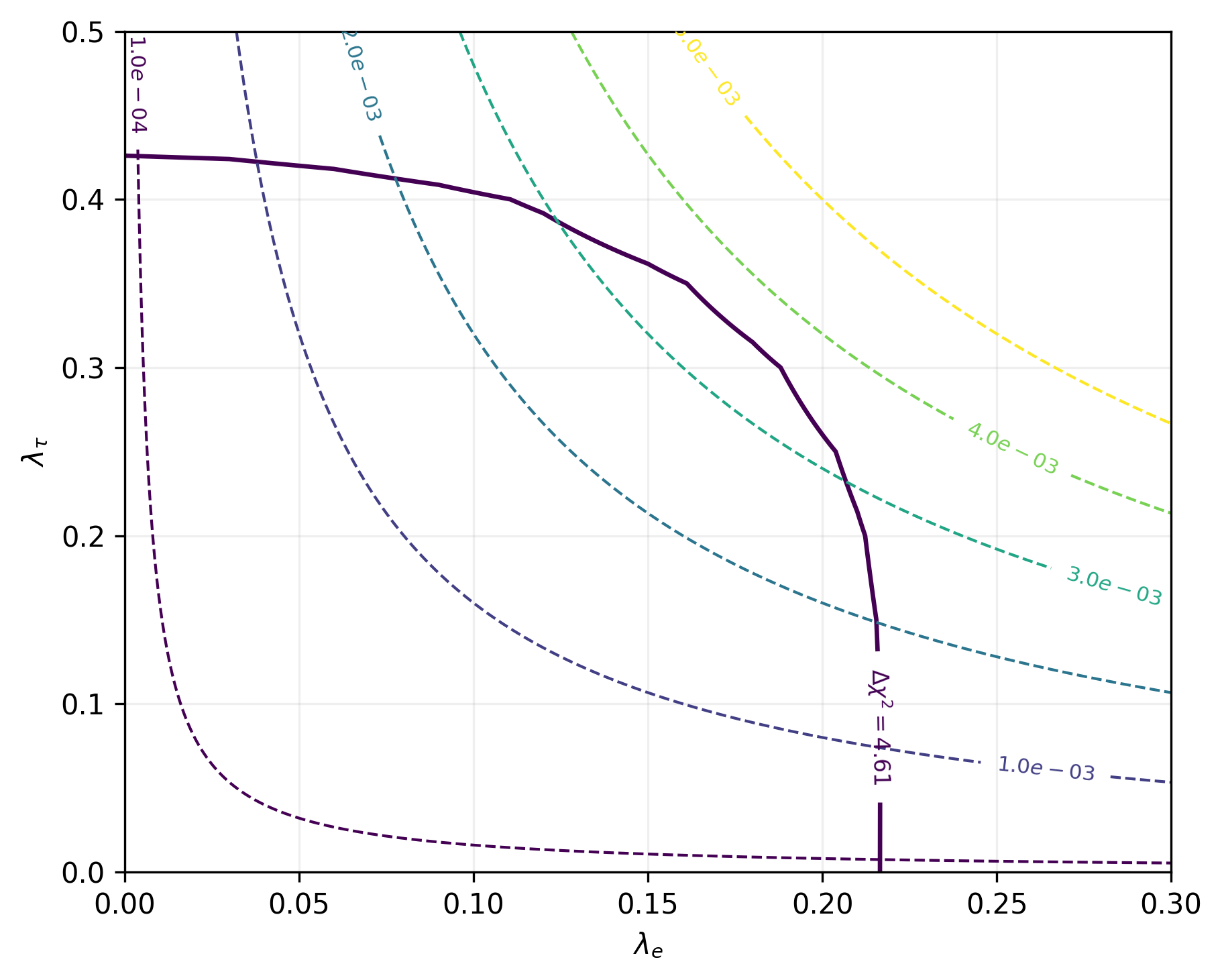}
    \hfill
    \includegraphics[scale=0.5]
        {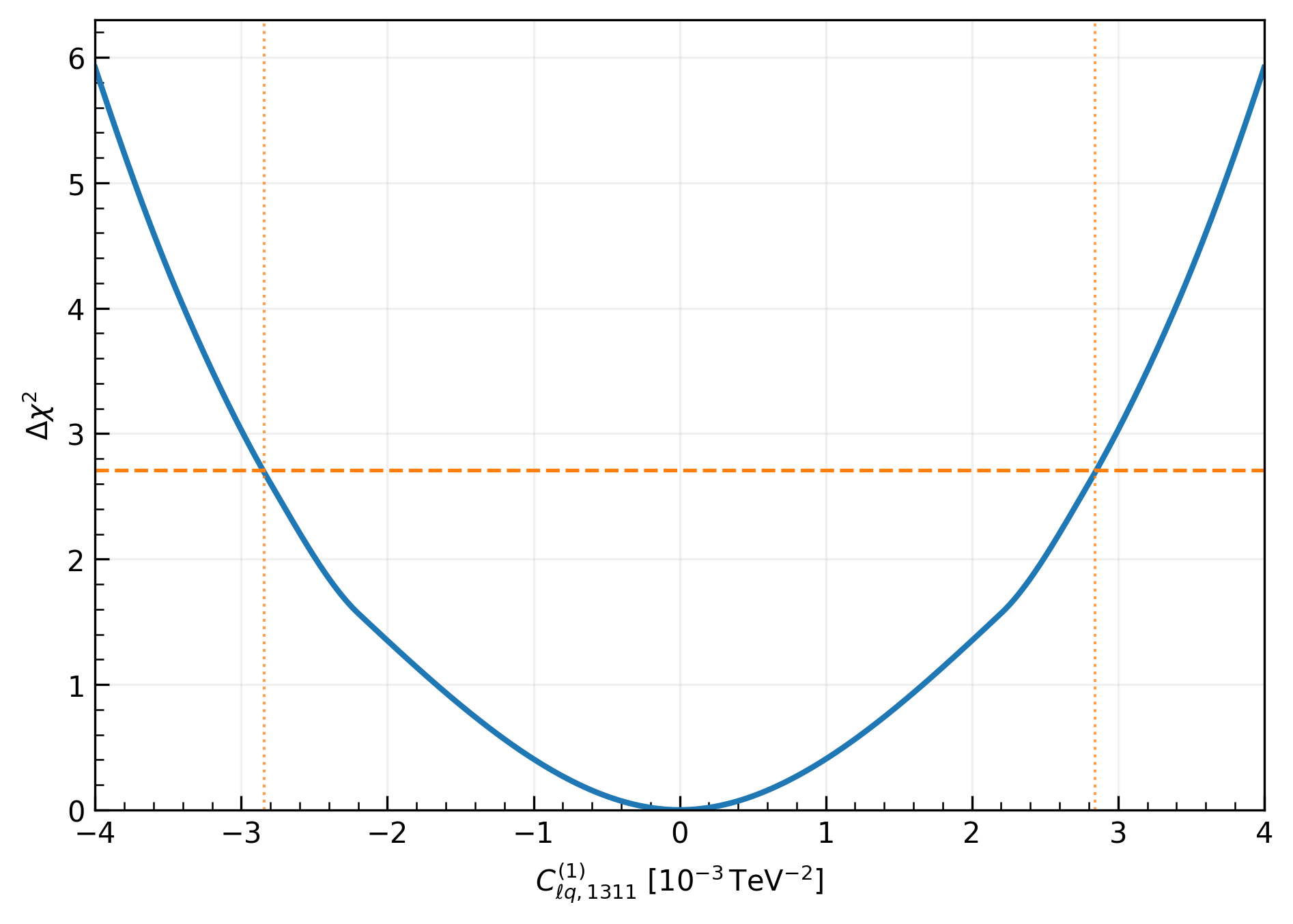}
    \caption{
      Left: likelihood in the $(\lambda_e,\lambda_\tau)$ plane for  $M_{\omega_1}=2~\mathrm{TeV}$. The dashed contours show constant values of $C_{\ell q,1311}^{(1)}$, while solid contour delimits the allowed region at $90\%$ C.L. Right: profile likelihood for the target coefficient $C_{\ell q,1311}^{(1)}$.
    }
    \label{fig:minimal_uv_results}
\end{figure*}

Combining all these constraints, we obtain, we obtain the two-dimensional $90\%$ C.L. allowed region in the plane of the couplings $\lambda_e$ and $\lambda_\tau$ in the left panel of
Fig.~\ref{fig:minimal_uv_results}. We also display the value of the WC target. As can be seen in the plot, the maximum values are around $(3-4)\times10^{-3}\;\rm{TeV}^{-2}$. 

To better determine the largest coefficient compatible with the data, we define
the fixed-target profile
\begin{equation}
    \chi^2(C)
    =
    \min_{\lambda_e,\lambda_\tau}
    \left\{
      \chi^2_{\mathrm{tot}}(\lambda_e,\lambda_\tau)
      \;\middle|\;
      C=
      \frac{\lambda_e^\ast\lambda_\tau}
           {4M_{\omega_1}^{2}}
    \right\}.
    \label{eq:uv_profile_definition}
\end{equation}
Operationally, at each fixed value of $C$ one coupling is eliminated using the matching condition, while the remaining coupling is varied to minimize the complete likelihood. 

The result is displayed in the right panel of
Fig.~\ref{fig:minimal_uv_results}. The profile is symmetric to excellent
accuracy and its minimum lies at the SM value. At $90\%$ C.L. we obtain
\begin{equation}
    \left|C^{(1)}_{\substack {\ell q \\ 1311}}\right|
    <
    2.9\times10^{-3}
    \ \mathrm{TeV}^{-2}.
    \label{eq:minimal_uv_bound}
\end{equation}
This corresponds to a bound on the product
\begin{equation}
    \left|\lambda_e\lambda_\tau\right|
    \lesssim
    4.6\times10^{-2}
    \left(\frac{M_{\omega_1}}{2~\mathrm{TeV}}\right)^2.
    \label{eq:minimal_uv_mass_scaling}
\end{equation}

It is immediate to notice that this result is substantially stronger than the tau-only one-coefficient
bound of Sec.~\ref{sec:eft_tau},
\begin{equation}
    \left|C_{\ell q,1311}^{(1)}\right|_{\tau\text{-only}}
    \lesssim
    8.1\times10^{-3}\ \mathrm{TeV}^{-2}.
    \label{eq:tau_only_comparison_uv}
\end{equation}
Thus, the maximal coefficient in the UV realization is smaller by a factor of approximately $2.8$. This result suggests that the tau-decay constraint has not been saturated. We found that, near the $90\%$ crossing, the increase in the likelihood (relative to its minimum) is dominated by the global fit from \texttt{smelli}, followed by $D$ mixing, while the direct tau-meson contribution is quite small. We obtained approximately
\begin{equation}
    \Delta\chi^2_{\mathrm{global}}
    :
    \Delta\chi^2_{D\text{-mix}}
    :
    \Delta\chi^2_{\tau\text{-mesons}}
    \simeq
    2.1:0.57:0.05.
    \label{eq:minimal_uv_chi2_decomposition}
\end{equation}

One of the main results from this exercise is the importance of taking into account D-mixing, which is a direct consequence of the one-loop matching we performed. 

\section{Summary and conclusions}
\label{sec:conclusions}

Flavour-changing neutrino non-standard interactions are generally difficult to realize in weakly coupled extensions of the SM above the electroweak scale. Gauge invariance typically relates the neutrino interaction to processes involving charged leptons, which are often subject to stringent flavour constraints~\cite{Davidson:2019iqh}. For this reason, recent analyses have generally focused only on flavour-conserving directions~\cite{Cherchiglia:2023aqp,Freitas:2025bgg}. Nevertheless, a comprehensive SMEFT analysis for DUNE~\cite{Kopp:2025ffx} identified the flavour-changing $\epsilon_{e\tau}$ direction as one of the cases for which the experiment could provide a significant improvement.

In this work, we have revisited this apparent window by confronting the DUNE-sensitive SMEFT direction with charged-lepton-flavour-violating tau decays. Within the semileptonic vector subspace spanned by $C_{\ell q,1311}^{(1)}$, $C_{\ell u,1311}$ and $C_{\ell d,1311}$, we showed that the direction probed by the NSI $\epsilon_{e\tau}$ is precisely the same as the isoscalar-vector combination entering the decay $\tau\to e\omega$. Consequently, a cancellation that eventually evades the $\tau\to e\omega$ bound will also suppress the corresponding  $\epsilon_{e\tau}$ direction. 

We explicitly evaluated the bounds coming from leptonic and semileptonic tau decays, showing that they, in general, will provide constraints one order of magnitude stronger than the projected DUNE's sensitivity. We have further investigated a minimal UV realization in relation to the tau decays bounds, showing that the simplest scenario, involving leptoquarks, is more strongly constrained by other observables such as $D^0$--$\bar D^0$ mixing, before the tau-decay bound is saturated. Therefore, UV completions capable of producing an $\epsilon_{e\tau}$ signal observable at DUNE are highly constrained.

We emphasize that our result is not a general no-go theorem for flavour-changing NSI. For instance, scenarios involving light degrees of freedom could in principle evade some of the bounds~\cite{Farzan:2015hkd,Farzan:2016wym}. It does show, however, that the apparently promising $\epsilon_{e\tau}$ direction shares the same SMEFT structure as some semileptonic tau decays, making it difficult to disentangle the charged-lepton-flavour constraints from the projected DUNE sensitivity.

\section*{Acknowledgements}
I thank Marceli Aquino for encouraging me to undertake this project. I thank Javier Fuentes-Martin, José Santiago, and the participants of HEFT-2026 for very fruitful discussions. I acknowledge support from the National Council for Scientific and Technological Development – CNPq, project 446121/2024-0. 

\bibliographystyle{JHEP} 
\bibliography{main}

@article{DUNE:2020ypp,
    author = "Abi, Babak and others",
    collaboration = "DUNE",
    title = "{Deep Underground Neutrino Experiment (DUNE), Far Detector Technical Design Report, Volume II: DUNE Physics}",
    eprint = "2002.03005",
    archivePrefix = "arXiv",
    primaryClass = "hep-ex",
    reportNumber = "FERMILAB-PUB-20-025-ND, FERMILAB-DESIGN-2020-02",
    month = "2",
    year = "2020"
}

@article{Farzan:2017xzy,
    author = "Farzan, Y. and Tortola, M.",
    title = "{Neutrino oscillations and Non-Standard Interactions}",
    eprint = "1710.09360",
    archivePrefix = "arXiv",
    primaryClass = "hep-ph",
    doi = "10.3389/fphy.2018.00010",
    journal = "Front. in Phys.",
    volume = "6",
    pages = "10",
    year = "2018"
}

@article{Bischer:2019ttk,
    author = "Bischer, Ingolf and Rodejohann, Werner",
    title = "{General neutrino interactions from an effective field theory perspective}",
    eprint = "1905.08699",
    archivePrefix = "arXiv",
    primaryClass = "hep-ph",
    doi = "10.1016/j.nuclphysb.2019.114746",
    journal = "Nucl. Phys. B",
    volume = "947",
    pages = "114746",
    year = "2019"
}

@article{Bergmann:1999pk,
    author = "Bergmann, Sven and Grossman, Yuval and Pierce, Damien M.",
    title = "{Can lepton flavor violating interactions explain the atmospheric neutrino problem?}",
    eprint = "hep-ph/9909390",
    archivePrefix = "arXiv",
    reportNumber = "SLAC-PUB-8244, WIS-99-31-DPP",
    doi = "10.1103/PhysRevD.61.053005",
    journal = "Phys. Rev. D",
    volume = "61",
    pages = "053005",
    year = "2000"
}

@article{Antusch:2008tz,
    author = "Antusch, Stefan and Baumann, Jochen P. and Fernandez-Martinez, Enrique",
    title = "{Non-Standard Neutrino Interactions with Matter from Physics Beyond the Standard Model}",
    eprint = "0807.1003",
    archivePrefix = "arXiv",
    primaryClass = "hep-ph",
    reportNumber = "MPP-2008-74",
    doi = "10.1016/j.nuclphysb.2008.11.018",
    journal = "Nucl. Phys. B",
    volume = "810",
    pages = "369--388",
    year = "2009"
}

@article{Gavela:2008ra,
    author = "Gavela, M. B. and Hernandez, D. and Ota, T. and Winter, W.",
    title = "{Large gauge invariant non-standard neutrino interactions}",
    eprint = "0809.3451",
    archivePrefix = "arXiv",
    primaryClass = "hep-ph",
    reportNumber = "FTUAM-08-15, IFT-UAM-CSIC-08-53, LPT-ORSAY-08-75",
    doi = "10.1103/PhysRevD.79.013007",
    journal = "Phys. Rev. D",
    volume = "79",
    pages = "013007",
    year = "2009"
}

@article{Meloni:2009cg,
    author = "Meloni, Davide and Ohlsson, Tommy and Winter, Walter and Zhang, He",
    title = "{Non-standard interactions versus non-unitary lepton flavor mixing at a neutrino factory}",
    eprint = "0912.2735",
    archivePrefix = "arXiv",
    primaryClass = "hep-ph",
    reportNumber = "NORDITA-2009-79, IDS-NF-014",
    doi = "10.1007/JHEP04(2010)041",
    journal = "JHEP",
    volume = "04",
    pages = "041",
    year = "2010"
}

@article{Altmannshofer:2018xyo,
    author = "Altmannshofer, Wolfgang and Tammaro, Michele and Zupan, Jure",
    title = "{Non-standard neutrino interactions and low energy experiments}",
    eprint = "1812.02778",
    archivePrefix = "arXiv",
    primaryClass = "hep-ph",
    doi = "10.1007/JHEP11(2021)113",
    journal = "JHEP",
    volume = "09",
    pages = "083",
    year = "2019",
    note = "[Erratum: JHEP 11, 113 (2021)]"
}

@article{Falkowski:2019kfn,
    author = "Falkowski, Adam and Gonz\'alez-Alonso, Mart\'\i{}n and Tabrizi, Zahra",
    title = "{Consistent QFT description of non-standard neutrino interactions}",
    eprint = "1910.02971",
    archivePrefix = "arXiv",
    primaryClass = "hep-ph",
    reportNumber = "IFIC/19-39, FTUV/19-1007, LPT-Orsay-19-35",
    doi = "10.1007/JHEP11(2020)048",
    journal = "JHEP",
    volume = "11",
    pages = "048",
    year = "2020"
}

@article{Falkowski:2019xoe,
    author = "Falkowski, Adam and Gonz\'alez-Alonso, Mart\'\i{}n and Tabrizi, Zahra",
    title = "{Reactor neutrino oscillations as constraints on Effective Field Theory}",
    eprint = "1901.04553",
    archivePrefix = "arXiv",
    primaryClass = "hep-ph",
    reportNumber = "CERN-TH-2019-002, LPT Orsay 19-02, CERN-TH-2019-002 LPT Orsay 19-02",
    doi = "10.1007/JHEP05(2019)173",
    journal = "JHEP",
    volume = "05",
    pages = "173",
    year = "2019"
}

@article{Babu:2019mfe,
    author = "Babu, K. S. and Dev, P. S. Bhupal and Jana, Sudip and Thapa, Anil",
    title = "{Non-Standard Interactions in Radiative Neutrino Mass Models}",
    eprint = "1907.09498",
    archivePrefix = "arXiv",
    primaryClass = "hep-ph",
    reportNumber = "FERMILAB-PUB-19-304-T, OSU-HEP-19-04",
    doi = "10.1007/JHEP03(2020)006",
    journal = "JHEP",
    volume = "03",
    pages = "006",
    year = "2020"
}

@article{Davidson:2019iqh,
    author = "Davidson, Sacha and Gorbahn, Martin",
    title = "{Charged lepton flavor change and nonstandard neutrino interactions}",
    eprint = "1909.07406",
    archivePrefix = "arXiv",
    primaryClass = "hep-ph",
    doi = "10.1103/PhysRevD.101.015010",
    journal = "Phys. Rev. D",
    volume = "101",
    number = "1",
    pages = "015010",
    year = "2020"
}

@article{Terol-Calvo:2019vck,
    author = "Terol-Calvo, Jorge and T\'ortola, Mariam and Vicente, Avelino",
    title = "{High-energy constraints from low-energy neutrino nonstandard interactions}",
    eprint = "1912.09131",
    archivePrefix = "arXiv",
    primaryClass = "hep-ph",
    reportNumber = "IFIC/19-60",
    doi = "10.1103/PhysRevD.101.095010",
    journal = "Phys. Rev. D",
    volume = "101",
    number = "9",
    pages = "095010",
    year = "2020"
}

@article{Babu:2020nna,
    author = "Babu, K. S. and Gon\c{c}alves, Dorival and Jana, Sudip and Machado, Pedro A. N.",
    title = "{Neutrino Non-Standard Interactions: Complementarity Between LHC and Oscillation Experiments}",
    eprint = "2003.03383",
    archivePrefix = "arXiv",
    primaryClass = "hep-ph",
    reportNumber = "FERMILAB-PUB-20-096-T, OSU-HEP-20-02",
    doi = "10.1016/j.physletb.2021.136131",
    journal = "Phys. Lett. B",
    volume = "815",
    pages = "136131",
    year = "2021"
}

@article{Du:2020dwr,
    author = "Du, Yong and Li, Hao-Lin and Tang, Jian and Vihonen, Sampsa and Yu, Jiang-Hao",
    title = "{Non-standard interactions in SMEFT confronted with terrestrial neutrino experiments}",
    eprint = "2011.14292",
    archivePrefix = "arXiv",
    primaryClass = "hep-ph",
    reportNumber = "ACFI-T20-15",
    doi = "10.1007/JHEP03(2021)019",
    journal = "JHEP",
    volume = "03",
    pages = "019",
    year = "2021"
}

@article{Falkowski:2021bkq,
    author = "Falkowski, Adam and Gonz\'alez-Alonso, Mart\'\i{}n and Kopp, Joachim and Soreq, Yotam and Tabrizi, Zahra",
    title = "{EFT at FASER\ensuremath{\nu}}",
    eprint = "2105.12136",
    archivePrefix = "arXiv",
    primaryClass = "hep-ph",
    doi = "10.1007/JHEP10(2021)086",
    journal = "JHEP",
    volume = "10",
    pages = "086",
    year = "2021"
}

@article{Du:2021rdg,
    author = "Du, Yong and Li, Hao-Lin and Tang, Jian and Vihonen, Sampsa and Yu, Jiang-Hao",
    title = "{Exploring SMEFT induced nonstandard interactions: From COHERENT to neutrino oscillations}",
    eprint = "2106.15800",
    archivePrefix = "arXiv",
    primaryClass = "hep-ph",
    doi = "10.1103/PhysRevD.105.075022",
    journal = "Phys. Rev. D",
    volume = "105",
    number = "7",
    pages = "075022",
    year = "2022"
}

@article{Breso-Pla:2023tnz,
    author = "Bres\'o-Pla, V\'\i{}ctor and Falkowski, Adam and Gonz\'alez-Alonso, Mart\'\i{}n and Mons\'alvez-Pozo, Kevin",
    title = "{EFT analysis of New Physics at COHERENT}",
    eprint = "2301.07036",
    archivePrefix = "arXiv",
    primaryClass = "hep-ph",
    doi = "10.1007/JHEP05(2023)074",
    journal = "JHEP",
    volume = "05",
    pages = "074",
    year = "2023"
}

@article{deBlas:2017xtg,
    author = "de Blas, J. and Criado, J. C. and Perez-Victoria, M. and Santiago, J.",
    title = "{Effective description of general extensions of the Standard Model: the complete tree-level dictionary}",
    eprint = "1711.10391",
    archivePrefix = "arXiv",
    primaryClass = "hep-ph",
    reportNumber = "CERN-TH-2017-251",
    doi = "10.1007/JHEP03(2018)109",
    journal = "JHEP",
    volume = "03",
    pages = "109",
    year = "2018"
}

@article{Cherchiglia:2023aqp,
    author = "Cherchiglia, Adriano and Santiago, Jose",
    title = "{DUNE potential as a new physics probe}",
    eprint = "2309.15924",
    archivePrefix = "arXiv",
    primaryClass = "hep-ph",
    doi = "10.1007/JHEP03(2024)018",
    journal = "JHEP",
    volume = "03",
    pages = "018",
    year = "2024"
}

@article{DUNE:2020jqi,
    author = "Abi, B. and others",
    collaboration = "DUNE",
    title = "{Long-baseline neutrino oscillation physics potential of the DUNE experiment}",
    eprint = "2006.16043",
    archivePrefix = "arXiv",
    primaryClass = "hep-ex",
    reportNumber = "FERMILAB-PUB-20-251-E-LBNF-ND-PIP2-SCD, PUB-20-251-E-LBNF-ND-PIP2-SCD",
    doi = "10.1140/epjc/s10052-020-08456-z",
    journal = "Eur. Phys. J. C",
    volume = "80",
    number = "10",
    pages = "978",
    year = "2020"
}

@article{Kopp:2024yvh,
    author = "Kopp, Joachim and Rocco, Noemi and Tabrizi, Zahra",
    title = "{Unleashing the power of EFT in neutrino-nucleus scattering}",
    eprint = "2401.07902",
    archivePrefix = "arXiv",
    primaryClass = "hep-ph",
    reportNumber = "FERMILAB-PUB-24-0017-T",
    doi = "10.1007/JHEP08(2024)187",
    journal = "JHEP",
    volume = "08",
    pages = "187",
    year = "2024"
}

@article{Cherchiglia:2025ufn,
    author = "Cherchiglia, Adriano",
    title = "{Unveiling New Physics Models Through Meson Decays and Their Impact on Neutrino Experiments}",
    doi = "10.3390/universe11070225",
    journal = "Universe",
    volume = "11",
    number = "7",
    pages = "225",
    year = "2025"
}

@article{Kopp:2025ffx,
    author = "Kopp, Joachim and Tabrizi, Zahra and Urrea, Salvador",
    title = "{Effective field theory in long-baseline neutrino oscillation experiments}",
    eprint = "2509.21537",
    archivePrefix = "arXiv",
    primaryClass = "hep-ph",
    doi = "10.1007/JHEP02(2026)176",
    journal = "JHEP",
    volume = "02",
    pages = "176",
    year = "2026"
}

@article{Guedes:2023azv,
    author = "Guedes, Guilherme and Olgoso, Pablo and Santiago, Jos\'e",
    title = "{Towards the one loop IR/UV dictionary in the SMEFT: One loop generated operators from new scalars and fermions}",
    eprint = "2303.16965",
    archivePrefix = "arXiv",
    primaryClass = "hep-ph",
    reportNumber = "DESY-23-040",
    doi = "10.21468/SciPostPhys.15.4.143",
    journal = "SciPost Phys.",
    volume = "15",
    number = "4",
    pages = "143",
    year = "2023"
}

@article{Guedes:2024vuf,
    author = "Guedes, Guilherme and Olgoso, Pablo",
    title = "{From the EFT to the UV: the complete SMEFT one-loop dictionary}",
    eprint = "2412.14253",
    archivePrefix = "arXiv",
    primaryClass = "hep-ph",
    reportNumber = "DESY-24-2",
    doi = "10.21468/SciPostPhys.20.3.074",
    journal = "SciPost Phys.",
    volume = "20",
    pages = "074",
    year = "2026"
}

@article{Straub:2018kue,
    author = "Straub, David M.",
    title = "{flavio: a Python package for flavour and precision phenomenology in the Standard Model and beyond}",
    eprint = "1810.08132",
    archivePrefix = "arXiv",
    primaryClass = "hep-ph",
    month = "10",
    year = "2018"
}

@article{Aebischer:2018bkb,
    author = "Aebischer, Jason and Kumar, Jacky and Straub, David M.",
    title = "{Wilson: a Python package for the running and matching of Wilson coefficients above and below the electroweak scale}",
    eprint = "1804.05033",
    archivePrefix = "arXiv",
    primaryClass = "hep-ph",
    doi = "10.1140/epjc/s10052-018-6492-7",
    journal = "Eur. Phys. J. C",
    volume = "78",
    number = "12",
    pages = "1026",
    year = "2018"
}

@article{Aebischer:2018iyb,
    author = "Aebischer, Jason and Kumar, Jacky and Stangl, Peter and Straub, David M.",
    title = "{A Global Likelihood for Precision Constraints and Flavour Anomalies}",
    eprint = "1810.07698",
    archivePrefix = "arXiv",
    primaryClass = "hep-ph",
    doi = "10.1140/epjc/s10052-019-6977-z",
    journal = "Eur. Phys. J. C",
    volume = "79",
    number = "6",
    pages = "509",
    year = "2019"
}

@article{Fernandez-Martinez:2024bxg,
    author = "Fern{\'a}ndez-Mart{\'\i}nez, Enrique and Marcano, Xabier and Naredo-Tuero, Daniel",
    title = "{Global lepton flavour violating constraints on new physics}",
    eprint = "2403.09772",
    archivePrefix = "arXiv",
    primaryClass = "hep-ph",
    reportNumber = "IFT-UAM/CSIC-24-39",
    doi = "10.1140/epjc/s10052-024-12973-6",
    journal = "Eur. Phys. J. C",
    volume = "84",
    number = "7",
    pages = "666",
    year = "2024"
}

@article{Wolfenstein:1977ue,
    author = "Wolfenstein, L.",
    title = "{Neutrino Oscillations in Matter}",
    reportNumber = "COO-3066-102",
    doi = "10.1103/PhysRevD.17.2369",
    journal = "Phys. Rev. D",
    volume = "17",
    pages = "2369--2374",
    year = "1978"
}

@article{Esteban:2018ppq,
    author = "Esteban, Ivan and Gonzalez-Garcia, M. C. and Maltoni, Michele and Martinez-Soler, Ivan and Salvado, Jordi",
    title = "{Updated constraints on non-standard interactions from global analysis of oscillation data}",
    eprint = "1805.04530",
    archivePrefix = "arXiv",
    primaryClass = "hep-ph",
    reportNumber = "IFT-UAM/CSIC-18-049, YITP-SB-18-11, IFT-UAM-CSIC-18-049",
    doi = "10.1007/JHEP08(2018)180",
    journal = "JHEP",
    volume = "08",
    pages = "180",
    year = "2018",
    note = "[Addendum: JHEP 12, 152 (2020)]"
}

@article{DUNE:2020fgq,
    author = "Abi, B. and others",
    collaboration = "DUNE",
    title = "{Prospects for beyond the Standard Model physics searches at the Deep Underground Neutrino Experiment}",
    eprint = "2008.12769",
    archivePrefix = "arXiv",
    primaryClass = "hep-ex",
    reportNumber = "FERMILAB-PUB-20-459-LBNF-ND, FERMILAB-PUB-20-459-LBNF-ND",
    doi = "10.1140/epjc/s10052-021-09007-w",
    journal = "Eur. Phys. J. C",
    volume = "81",
    number = "4",
    pages = "322",
    year = "2021"
}

@article{Arguelles:2019xgp,
    author = {Arg{\"u}elles, C. A. and others},
    title = "{New opportunities at the next-generation neutrino experiments I: BSM neutrino physics and dark matter}",
    eprint = "1907.08311",
    archivePrefix = "arXiv",
    primaryClass = "hep-ph",
    reportNumber = "FERMILAB-FN-1079-T",
    doi = "10.1088/1361-6633/ab9d12",
    journal = "Rept. Prog. Phys.",
    volume = "83",
    number = "12",
    pages = "124201",
    year = "2020"
}

@article{Mikheyev:1985zog,
    author = "Mikheyev, S. P. and Smirnov, A. Yu.",
    title = "{Resonance Amplification of Oscillations in Matter and Spectroscopy of Solar Neutrinos}",
    journal = "Sov. J. Nucl. Phys.",
    volume = "42",
    pages = "913--917",
    year = "1985"
}

@article{Gonzalez-Alonso:2026sgl,
    author = "Gonz{\'a}lez-Alonso, Mart{\'\i}n and Palavri{\'c}, Ajdin and Prakash, Suraj",
    title = "{EFT for Neutrino Oscillations: Theory Developments and Application to JUNO}",
    eprint = "2606.11298",
    archivePrefix = "arXiv",
    primaryClass = "hep-ph",
    month = "6",
    year = "2026"
}

@article{Coloma:2023ixt,
    author = "Coloma, Pilar and Gonzalez-Garcia, M. C. and Maltoni, Michele and Pinheiro, Jo{\~a}o Paulo and Urrea, Salvador",
    title = "{Global constraints on non-standard neutrino interactions with quarks and electrons}",
    eprint = "2305.07698",
    archivePrefix = "arXiv",
    primaryClass = "hep-ph",
    reportNumber = "IFT-UAM/CSIC-23-47, IFIC/23-15, FTUV-23-0427.3710, YITP-SB-2023-05",
    doi = "10.1007/JHEP08(2023)032",
    journal = "JHEP",
    volume = "08",
    pages = "032",
    year = "2023"
}

@article{Coloma:2024ict,
    author = "Coloma, Pilar and Fern{\'a}ndez-Mart{\'\i}nez, Enrique and L{\'o}pez-Pav{\'o}n, Jacobo and Marcano, Xabier and Naredo-Tuero, Daniel and Urrea, Salvador",
    title = "{Improving the global SMEFT picture with bounds on neutrino NSI}",
    eprint = "2411.00090",
    archivePrefix = "arXiv",
    primaryClass = "hep-ph",
    reportNumber = "IFT-UAM/CSIC-24-151, FTUV-24-1025.8856",
    doi = "10.1007/JHEP02(2025)137",
    journal = "JHEP",
    volume = "02",
    pages = "137",
    year = "2025"
}

@article{Belle:2007cio,
    author = "Miyazaki, Y. and others",
    collaboration = "Belle",
    title = "{Search for lepton flavor violating tau- decays into l- eta, l- eta-prime and l- pi0}",
    eprint = "hep-ex/0703009",
    archivePrefix = "arXiv",
    reportNumber = "BELLE-PRERPINT-2007-13, KEK-PRERPINT-2006-78",
    doi = "10.1016/j.physletb.2007.03.027",
    journal = "Phys. Lett. B",
    volume = "648",
    pages = "341--350",
    year = "2007"
}

@article{Belle:2023ziz,
    author = "Tsuzuki, N. and others",
    collaboration = "Belle",
    title = "{Search for lepton-flavor-violating {\ensuremath{\tau}} decays into a lepton and a vector meson using the full Belle data sample}",
    eprint = "2301.03768",
    archivePrefix = "arXiv",
    primaryClass = "hep-ex",
    reportNumber = "Belle Preprint 2022-33; KEK Preprint 2022-46",
    doi = "10.1007/JHEP06(2023)118",
    journal = "JHEP",
    volume = "06",
    pages = "118",
    year = "2023"
}

@article{Allwicher:2022gkm,
    author = "Allwicher, Lukas and Faroughy, Darius A. and Jaffredo, Florentin and Sumensari, Olcyr and Wilsch, Felix",
    title = "{Drell-Yan tails beyond the Standard Model}",
    eprint = "2207.10714",
    archivePrefix = "arXiv",
    primaryClass = "hep-ph",
    doi = "10.1007/JHEP03(2023)064",
    journal = "JHEP",
    volume = "03",
    pages = "064",
    year = "2023"
}

@article{DiLuzio:2018zxy,
    author = "Di Luzio, Luca and Fuentes-Martin, Javier and Greljo, Admir and Nardecchia, Marco and Renner, Sophie",
    title = "{Maximal Flavour Violation: a Cabibbo mechanism for leptoquarks}",
    eprint = "1808.00942",
    archivePrefix = "arXiv",
    primaryClass = "hep-ph",
    reportNumber = "IPPP/18/69, MITP/18-072, CERN-TH-2018-182",
    doi = "10.1007/JHEP11(2018)081",
    journal = "JHEP",
    volume = "11",
    pages = "081",
    year = "2018"
}

@article{Freitas:2025bgg,
    author = "Freitas, Ayres and Low, Matthew",
    title = "{Non-standard neutrino interactions at neutrino experiments and colliders}",
    eprint = "2505.01401",
    archivePrefix = "arXiv",
    primaryClass = "hep-ph",
    doi = "10.1007/JHEP10(2025)067",
    journal = "JHEP",
    volume = "10",
    pages = "067",
    year = "2025"
}

@article{Farzan:2016wym,
    author = "Farzan, Yasaman and Heeck, Julian",
    title = "{Neutrinophilic nonstandard interactions}",
    eprint = "1607.07616",
    archivePrefix = "arXiv",
    primaryClass = "hep-ph",
    reportNumber = "ULB-TH-16-12",
    doi = "10.1103/PhysRevD.94.053010",
    journal = "Phys. Rev. D",
    volume = "94",
    number = "5",
    pages = "053010",
    year = "2016"
}

@article{Farzan:2015hkd,
    author = "Farzan, Yasaman and Shoemaker, Ian M.",
    title = "{Lepton Flavor Violating Non-Standard Interactions via Light Mediators}",
    eprint = "1512.09147",
    archivePrefix = "arXiv",
    primaryClass = "hep-ph",
    doi = "10.1007/JHEP07(2016)033",
    journal = "JHEP",
    volume = "07",
    pages = "033",
    year = "2016"
}

\end{document}